# Breaching the privacy of connected vehicles network

Vladimir Kaplun and Michael Segal
Ben-Gurion University of the Negev
segal@bgu.ac.il

***Abstract*— Connected Vehicles network is designed to provide a secure and private method for drivers to use the most efficiently the roads in certain area. When dealing with the scenario of car to access points connectivity (Wi-Fi, 3G, LTE), the vehicles are connected by central authority like cloud. Thus, they can be monitored and analyzed by the cloud which can provide certain services to the driver, i.e. usage based insurance (UBI), entertainment services, navigation etc.**

**The main objective of this work is to show that by analyzing the information about a driver which is provided to the usage based insurance companies, it is possible to get additional private data, even if the basic data in first look, seems not so harmful. In this work, we present an analysis of a novel approach for reconstructing driver's path from other driving attributes, such as cornering events, average speed and total driving time. We show that, in some cases, it is possible to reconstruct the driver's path, while not knowing the target point of the trip.**[1]

## I. INTRODUCTION

The Internet of Things (IoT) is a new trend in our information and communication process stemming from the evolution of the Internet. There are many use cases in which IoT technologies are explored like smart phones, watches, electrical devices and more. One of the most interesting fields that we plan to investigate is a cloud monitored vehicles, in which trip information (location and time) is generated by each vehicle and stored in cloud's database. The information that is stored in the cloud's database can be very valuable for third party companies. Their interest can be due to many utilities that the data gathered from the vehicle can provide, e.g. tracking, learning patterns, providing usage based insurance, learning statistics about road conditions and more. Therefore, it is important to ensure a privacy of the user in a process of queries which are performed by the various third party companies.

In this work we will focus on usage-based-insurance (UBI). A usage-based-insurance is an automobile insurance where the insurer uses data on driving behavior to set the premium offered to each policyholder. The premiums are adjusted so as to reflect the individual driver risk profiles constructed by the insurer. In order to calculate the risk of each driver properly, the insurance company has to know several driving attributes e.g. total driving time, cornering, and average speed. Commercial UBI programs are available on the market today are mainly based on information extracted from the car's on-board-diagnostics (OBD) system, or from externally installed hardware components, referred to as black-boxes or aftermarket devices. Another method for measuring cornering and other attributes for the UBI revenues are smartphone-based insurance telematics applications, aiming to avoid the logistic and monetary costs associated with on-board-diagnostics or black-box dependent solutions.

The aim of this thesis is to identify whether the privacy of the users can be compromised by the usage-based-insurance companies. The privacy breach can be reached by getting basic information about specific user and by using the algorithms that we conducted. Thus, throughout the thesis we will show that it is possible to find user's path by knowing some attributes that the UBI companies gather from the driver in order to assess the level of each user's risk.

This work is organized as follows: Section II presents the problem definition and describes the model used in this research. Literature survey and previous work description can be found in Section III. Section IV describes our algorithms for revealing driver's path. In Section V we show extended simulation results and finally Section VI concludes our work.

## II. PRELIMINARIES AND MODEL

This section provides a description of the model used in this research and the required notations. In addition, it includes a definition of the problem studied.

### A. Model

The routing algorithm is assumed to be an on-demand algorithm, i.e., a path between a source node and a destination node is set up only when a request is made.

We start by listing the graph theory notations which are used in this work.

Connected vehicles network is well presented using graph theory, while the roads are presented as a collection of directed edges and the intersections are presented as a collection of vertices. Intersections are defined as the junction at-grade of two or more roads meeting or crossing. Furthermore, we also define intersections as turning events greater than 60°. Let some directed graph $G$ represent a road map inside a defined area. We let $V(G)$, $E(G)$ to represent the sets of vertices (intersections) and edges (roads), respectively of $G$, where $|V(G)| = n$. A directed edge $e_{v_i,v_j} = (v_i, v_j) \in E$ exists if a vehicle can reach $v_j$ from $v_i$ in 1 hop path. The use of directed graph comes from constraints on the direction of driving in the physical world. If there is a directed edge which connects vertex $v_i$ to $v_j$, vertex $v_j$ is considered as a successor of $v_i$. If there is a successor for the vertex, it is possible for the driver to drive to the next intersection.

In addition, we define a simple path as a set of disjoint vertices $[v_i, \dots, v_j]$, which are connected by edges, while one can reach the last vertex from the first vertex using the directed edges. We define the length of the path $Path$ as a number of vertices that path contains, and denote it as $|Path|$.

The use of weighted directed graph comes from the legal and physical constraints of each road. There are 3 weights for each of the edges $e_{v_i,v_j} \in E$ in graph $G$:

- We define $W(e_{v_i,v_j})$ as the maximum legal speed which is allowed by law in the road between $v_i$ to $v_j$, and define $W([v_i, v_j, \dots, v_l, v_k]) = \frac{W(e_{v_i,v_j}) + \cdots + W(e_{v_l,v_k})}{|[v_i, v_j, \dots, v_l, v_k]| - 1}$ which is average speed in $[v_i, v_j, \dots, v_l, v_k]$ $path$.
- We use $d(e_{v_i,v_j}) = |e_{v_i,v_j}|$ to denote the road distance of $e_{v_i,v_j}$ which connects vertices $v_i, v_j \in V(G)$. We assume that $d(v_i, v_j)$ is limited by $min_{road} \leq d(e_{v_i,v_j}) \leq max_{road}$. In order to denote the distance of a certain path, we use the following notations:
- We define $d([v_i, v_j, \dots, v_k])$ as a distance of the $[v_i, v_j, \dots, v_k]$ path.
- We define $d([v_i, v_j, \dots, v_k], v_a)$ as a distance of the $[v_i, v_j, \dots, v_k, v_a]$ path, while $v_a$ is disjoint from $v_i, v_j, \dots, v_k$ vertices.

There are 2 limits on the distance which we define as follows:

- We define $X_{turn}$, a distance between 2 turns to the same direction:
$$min_{road} \leq X_{turn} \leq max_{turn}$$
- Let's define $max_{dist}$ as a maximum distance between the starting vertex and the possible turn. In general, it is bounded by $|E| \cdot max_{road}$. However, in our algorithms we can limit it with maximum legal speed and time difference between the starting vertex and possible turning event.

In order to map the popularity of the roads in certain area, we use $P(e_{v_i,v_j})$ to denote the road popularity which connects vertices $v_i, v_j \in V(G)$. If we want to denote the average popularity of certain directed path, we would use $P([v_i, \dots, v_j])$.

Furthermore, in order to have exact definition of a cornering event we define a turning angle as at least 60°, similarly to [1]. Therefore, if an angle between the previous road direction and the current road direction is between 0-59°, it would be considered as straight driving, and if the angle is between 60°-180°, it would be considered as a turning event. When dealing with calculation of the closest edge to a specific GPS coordinate, we introduce the definition of $harvesian_distance(p, e)$ which is a straight segment distance between GPS point $p$ and edge (road) $e$.

In addition to the edge weights, it is important to define the direction of each edge (road). Thus, we introduce the $road\ direction$. We will use $< v_i, v_j >$ to define the 2 dimensional $road\ direction$ vector of $e_{v_i,v_j}$, while each $road\ direction$ is relative to the north of the planet. Furthermore, we later on will use the notion of $Previous_{road\ direction}$ vector, which is the vector of predecessor road. Predecessor road, is a road in which former vertex is the first vertex of the currently examined road.

Finally, in order to assess the number of turning events inside the certain path, we use the notation of $turn_num_{Path}$.

### B. Problem Definition

This work presents a study of the problem of **Breaching Drivers Privacy** by revealing driver's path while using basic driving information. The motivation is to find an efficient algorithm in a good computation time, which would hopefully reveal driver's path while the starting point is given. Specifically, we would like to maximize the revealed path distance, while minimizing the variance between the real driver's path and the revealed path. Furthermore, we would like to have the best revealed path which has the highest popularity among possible paths.

Since there are vast amount of possibilities for turning in a specific amount of seconds, it would be very challenging to find the correct path. Thus, the objective of this work is to determine how to use the given driver's attributes, and to find the influence of the road popularities on breaching driver's privacy.

## III. PREVIOUS WORK

This section reviews the previous works which were performed in the field of privacy and usage based insurance path mitigation. We first survey the works which defines the user privacy and ways to protect it. Afterwards, we examine some works which deal with usage based insurance and their threat to user's privacy.

### A. Privacy classification

In order to understand the threats that UBI possesses to the privacy of the drivers, it is important to classify and measure privacy levels. Thus, restraining the queries which the third party companies ask the database. Therefore, Xiaofeng et al. [2] suggests classifying the privacy levels by two parameters: universality and confidentiality. The privacy universality indicates how many people think their privacy is impaired when the information is disclosed. While, the privacy confidentiality indicates the importance of the privacy to the data owner and the degree of secrecy.

In addition, there are many other methods for classification of privacy levels such as using machine learning algorithms in order to compute the mutual information between the utility and privacy. The utility of a dataset is a measure of how useful a privatized dataset is to the dataset owner. Thus, by setting a privacy threshold, which defines the levels of privacy inside the dataset, the utility of the query to the dataset of the third party can be restricted. As a result, it can prevent privacy breach [3]. To the best of our knowledge, the path of the user is the most valuable private data of each user [4]. Thus, in our work we focus on inferring the path of the user which is considered to be the highest private data. This is because UBI companies can infer other private attributes from driver's path, like personal address, working address and the places that the user has visited. For example, it would be very dangerous for any politician to expose his daily pass, accordingly, exposing himself to unnecessary threat.

### B. Privacy Anonymization

The main threat to user's privacy is inferring an additional information about a driver from an information table release. The first model of privacy-preserving data publication was *k-anonymity* [5]. That model suggests to generalize the values of the attributes so that each of the released record becomes indistinguishable from at least *k-1* other records, when projected on those attributes. As a consequence, each individual may be linked to sets of records of size at least $k$ in the released anonymized table, whence privacy is protected to some extent. While *k-anonymity* refers only to single release of the table, protecting the private information from adversaries who examine the sequential release was studied in [6]. Wang et al. introduces the "lossy join" which generalizes the current release of the table column, so that the join with the previous release of the table column becomes lossy enough to disorient the attacker. Shmueli et al. [7] further investigates the notion of protecting the data in sequential releases and extends the study of continuous data publishing. In their study, they present 2 privacy attributes, *k-linkability* and *k-diversity*. The *k-linkability* mandates that even if an adversary combines information from all releases of the underlying table, he would not be able to link any selection of values of the attributes to less than $k$ distinct values of the sensitive attribute. The *k-diversity* demands that such an adversary would not be able to link any selection of values of the attributes with any sensitive value with probability greater than $1/k$. In order to achieve the requirements above, the paper [7] proposes "Cell-Generalization" method, in which each cell is generalized independently.

### C. Usage Based Insurance data aggregation methods

There are several methods for Usage Based Insurance to gain user's data. Such methods can gain small portion of drivers' attributes or even all them. One of those methods is vehicle telematics based program. In order to enter the program, a driver has to install telematics unit which in turn gains user's mileage, breaking habits, time of a day when the data was recorded and average speed. Furthermore, some telematics units aggregate cornering behavior log of the driver [8]. Given the privacy issues surrounding the geographic tracking of individuals, many solutions explicitly claim that the customer's GPS coordinates are not recorded. Privacy policies clearly state what information is collected, as well as the possibility of sharing the data with third-parties, using it for fraud prevention and research, or for compliance with the law [9]. Recent estimates predict that up to 30% of all vehicles in the United States, and 60% of all vehicles in the United Kingdom, will be insured through some type of insurance telematics program by the year 2020 [10].

Because of large costs related to installation, maintenance, and logistics which involved with telematics programs, another method for aggregating driver's attributes was presented in [11]–[14]. That method involves a smartphone-based insurance telematics applications. Currently, the commercial expansion of the UBI industry is held up by the process of acquiring data. On one hand, the use of smartphones for the collection of driving data is much simpler than telematics methods, due to the high percentage of drivers who own a smartphone. On the other hand, the vast amount of information that can be collected from the smartphone can infer driver's privacy [15].

## D. Cornering data

Despite efforts to improve the conditions of the road surface and the quality of the tires, skidding and rollover events still play a major role in many of today's car accidents. Moreover, statistics show that even though only three percent of all vehicle crashes involve a rollover, approximately 1/3 of all passenger deaths are related to rollover events [16]. As of yet, no safety system exists that can fully compensate for the dangers in turning events induced by excessive speeds or reckless driving. Thus, in order to perform a better risk analyses of the driver's driving skills, it is important to detect dangerous cornering events.

## E. Path Finding Algorithms

One of the most interesting privacy breach attacks, is an attack which gains path from some driver's attributes. Hunter at el. [117] presented an algorithm of reconstructing vehicle trajectories from sparse sequences of GPS points, for which the sampling interval ranges between 10 seconds and 2 minutes. The algorithm maps streaming GPS data in real-time, with a high throughput. They present an efficient Expectation Maximization algorithm to train the filter on new data without ground truth observations. Two of the common problems which occur when dealing with these GPS traces are the correct mapping of these observations to the road network, and the reconstruction of the trajectories of the vehicles from these traces. The main challenge is finding the right path among very high possible paths due to urban environment. The main disadvantage of that algorithm is despite of its success reconstructing driver's path, it needs points in the middle and the end of the path. In our work, we assume that we have only the starting point and the cornering events. Thus, we do not map the GPS traces.

Another interesting work which inferred driver's path from another attributes was performed by Dewri et al. [1]. In their study, they showed that the destinations of trips may also be determined without having to record GPS coordinates. In this paper, they studied the threat of location inference in vehicle telematics applications that collect driving habits data. Hence, developing an inference algorithm to demonstrate that inferring the destinations of driving trips is possible with access to simple features such as driving speed and distance traveled. The algorithm does fail in some cases, e.g. traffic jams. In order to work, it needs an ideal road and turning conditions. Thus, when there is a traffic jam or if the driver didn't take a turn in the right speed or pattern it wouldn't work. Furthermore, the researcher considers that every driver, always takes a shortest path to the destination. In our work, we do not make that assumption and our algorithm can work even if there is a traffic jam.

The work of Gao et al. [18] shows that drivers can be tracked by merely collecting their speed data and knowing their home location. To demonstrate the algorithm's real-world applicability, they evaluated its performance with datasets which represents suburban and urban areas. The algorithm predicted destinations with error within 250 meters for 14% traces and within 500 meters for 24% traces one dataset (254 traces). For a larger dataset (691 traces), they similarly predicted destinations with error within 250 and 500 meters for 13% and 26% of the traces respectively. Thus, showing that these insurance schemes enable a substantial breach of privacy. The percent of predicted endpoints within 250 meters of the actual endpoint also does not decrease with distance, with trips as long as 10.5 miles still having endpoints correctly predicted to within 250 meters. Unfortunately, the main assumption of the algorithm is that the speed is known at least in a rate of 2 samples in second, very high sampling rate, since there is sometimes a loss in GPS signal. In our work, we are basing our solutions on an average speed of the driver instead of continuous speed data.

If the attacker wishes to find user's path, he has to rely only on the information that the UBI companies need in order to assess the risk of the driver, while the combination of all of them can cause a privacy breach. As we mentioned beforehand, the attributes that are provided are starting point, cornering log file, and average speed. Cornering data is provided about the speed pattern when a driver performs a turn, left or right. Thus, when performed dangerously, it would cause much higher insurance payment. As shown in [19] and [9], left turn differs from right turn in some features like higher speed in left turn and different speed pattern. Hence, the detection of left turn, right turn is performed by matching training templates for these events with some test data. In our algorithms we assume that the cornering data is provided to us after the detection of left and right turn and the time when the turning event occurred.

# IV. UBI ATTACK ALGORITHM

Within this section, we present, describe and analyze a new approach for discovering driver's trajectory from attributes that are provided to the UBI companies.

## A. Mapping popularities in graph G

Before the use of the algorithm, it is important to map all of the road popularities. Thus, we propose $Mapping\ Popularities$ algorithm in order to perform this task.

The input of the $Mapping\ Popularities$ algorithm receives $G = (V, E)$, directed graph which represents the driving area. In addition, in order to compute popularity weight for each of the edges, we also need GPS log files, where each one of them encapsulates the GPS coordinates of specific road user (vehicle). The GPS log files are defined as $Pfile$ array. The GPS coordinates of each vehicle $i$ in $Pfile$ is formatted as follows, while $GPS_z$ represents GPS coordinates of the vehicle.In our algorithm we denote GPS data $j$ inside $Pfile\ i$ as $Pfile[i][j]$. The algorithm returns graph $G = (V, E)$ with adjacent popularity weight $P(e)$ for each edge $e$.

---

$\boldsymbol{Getting\ Popularities}\ (\boldsymbol{G}, \boldsymbol{PFile})$

$\boldsymbol{for}\ each\ e\ in\ E$:
  $P(e)\ = 0$
$\boldsymbol{for}\ i \leftarrow 1\ to\ i \leftarrow number\ of\ Pfiles$:
  $\boldsymbol{for}\ j \leftarrow 1\ to\ j \leftarrow\ Pfile[i].length$:
    $p =\ Pfile[i][j]$
    $d = \infty$
    $e_{closest} = null$
    $\boldsymbol{for}\ each\ e\ in\ E$:
      $\boldsymbol{if}\ harvesian_distance(p, e) < d$:
        $d = harvesian_distance(p, e)$
        $e_{closest} = e$
    $P(e_{closest}) += 1$
$\boldsymbol{return}\ G$

---

The computation time of $Getting\ Popularities$ algorithm is equal to $O(|E| \cdot |C| \cdot |N|)$, while $|E|$ represents a number of edges in the graph, $|C|$ is the number of files and $|N|$ represents a number of GPS coordinates in each file. Furthermore, if the number of GPS coordinates is very high, we can construct Voronoi Diagram for finding the closest edge to each GPS coordinate. Thus, we can take all of the GPS coordinates and build a Voronoi Diagram for them, which will consume computation time of $O(|C| \cdot |N| \log(|C| \cdot |N|))$. Afterwards, for each edge, we find which Voronoi cells it intersects. Therefore, if $k$ represents the maximum number of voronoi cells that can be intersected by single edge, the total computation time would be $O(|C| \cdot |N| \log(|C| \cdot |N|) + |E| \cdot k)$.

**B. Finding Straight Paths From a certain vertex**

In order to find driver's path, first, it is important to find the possible paths between the last vertex in the $path$ and possible turn event which will be used later on. This is the goal of $Straight\ Path\ Finder$ algorithm. In addition, we limit $Straight\ Path\ Finder$ algorithm to find paths with distance which is no greater than *Y* kilometers.

There are several functions which we use in the algorithm. The "append_vertex" function is used when we want to add a vertex to path. The "append_one_path" function is used when we want to add an entire path to the array. The "append_paths" is used when we want to add several paths to an array. In addition, the "add_vertex" function is used when we want temporarily add a certain vertex to the end of the path. Finally, the "add_paths" is used when we want temporarily add one path to another, while the second path is added in the end of the first path. Thus, in following code lines path will not contain the added vertex or path. In the input of the $Straight\ Path\ Finder$ algorithm there are attributes that assist the algorithm to find possible paths. The algorithm receives $\Delta t$ which is the time difference between the starting vertex and the turning event. Another attribute is a direction (left or right) that the driver made a turn to, which we will denote as $Direction\ of\ Turn$. In addition, we define $residential_{speed}$ as a speed in residential area.

Furthermore, the algorithm has to know what happened before it was called, e.g. previous path, and whether it is another recursion call, which complies when $From\ Turn$ Boolean is False. It gets $Path$ attribute which is a path until the current vertex, and $From\ Turn$ which is a Boolean that indicates whether the previous road was a turn event.

In the output of the algorithm, we get a set of paths which go straight, starting from the last vertex of the $Path$ until possible turning vertices under the constraints of $\Delta t$ and $Direction\ of\ Turn$.

---

$\boldsymbol{Straight\ Path\ Finder}$
$(\ \boldsymbol{\Delta t}\ , \boldsymbol{Direction\ of\ Turn}, \boldsymbol{Path}\ , \boldsymbol{From\ Turn})$

---

$B = \{\}$
$\boldsymbol{If}\ |Path| \geq 2$:
  $Previous_{road\ direction} = Get\ Direction\ from\ 2\ last\ vertices\ in\ Path$
$\boldsymbol{else}$: $Previous_{road\ direction} = null$
$Node = last\ vertex\ in\ Path$
$Successors\ = Find\ all\ Node's\ neighbors\ in\ G(V, E)$
$\boldsymbol{If}\ length(Successors) == 0$:
  $Mark\ Path\ as\ "closed"$
  $B.append_one_path(Path)$
  $\boldsymbol{return}\ B$
$\boldsymbol{else}$:
  $X = residential_{speed} * \Delta t$
  $\boldsymbol{for}\ i \leftarrow 1\ to\ i \leftarrow length(Successors)$:
    $\boldsymbol{If}\ Previous_{road\ direction}$ is not $null$:
      $Angle, turn =$ Find angle, turn between the previous and current direction
      $\boldsymbol{If}\ 180° > angle > 59°\ and\ turn == Direction\ of\ turn$:
        $\boldsymbol{If}\ From\ Turn\ ==\ False$:
          $Mark\ Path[length(Path) - 1]\ as\ "turn"$
          $B.append_one_path\ (Path.add_vertex("closed"))$
        $\boldsymbol{else}$:
          $B.append_{one_{path}}(Path.add_vertex(S\text{uccessors}[i]))$
      $\boldsymbol{If}\ X < W(Path, Successors[i]) * \Delta t$:
        $X\ =$

---

$W(Path, Successors[i]) * \Delta t$

24) ***else if*** $angle < 60°$ && $From\ Turn == False$:

25) ***If*** $d(Path, Successors[i]) < W(Path, Successors[i]) * \Delta t$:

26) $B.append_{one_{path}}(Path.add_vertex(Successors[i]))$

27) ***If*** $X < W(Path, Successors[i]) * \Delta t$:

28) $X = W(Path, Successors[i]) * \Delta t$

29) ***else***: $B.append_{one_{path}}(Path.add_vertex("closed")$

30) **else**: ***If*** $d(Path, Successors[i]) < W(Path, Successors[i]) * \Delta t$:

31) $B.append_one_path(Path.add_{vertex}(Successors[i])$

32) ***If*** $X < W(Path, Successors[i]) * \Delta t$:

33) $X = W(Path, Successors[i]) * \Delta t$

34) ***else***: $B.append_one_path(Path.add_{vertex}("closed"$

35) $A = h\ most\ popular\ opened\ paths\ in\ B$

36) $B =\ closed\ paths\ in\ B$

37) ***for*** $i$ ← 1 $to$ $i$ ← $Length[A]$:

38) $B.append_paths($***Straight Path Finder*** $(\Delta t, Direction\ of\ Turn, A[i], False))$

39) ***return*** $B$

The $Straight\ Path\ Finder$ algorithm defines an $B$ array, and $Previous_{road\ direction}$. The $B$ array, later on will encapsulate all of our possible paths until the possible turn. Whereas, the variable $Previous_{road\ direction}$ indicates the vector of the previous road. Consequently, in steps 7-10, the input $path$ is marked as "closed" and $B$ is retuned if there are no successors, otherwise the algorithm proceeds to the12-34 steps, which in turn, adds the path and their successors to $B$ array and assigns the longest possible distance to $X$ variable. The $X$ variable would store the maximum distance that can be reached with the boundary of $\Delta t$ and maximum legal speed, while the least maximum legal speed equals to $residential_{speed}$.

In steps 13-34, the algorithm examines each of the successors of the $Node$. It calculates the angle between the $Previous_{road\ direction}$, which was calculated in 2-4 steps, and the current road direction and determines whether the current road is considered as a turning event to the same direction as $Direction\ of\ Turn$. In that case, the successor is marked as a "turn", the p$ath$ and its successor are marked as "closed", and appended to $B$ array. In other cases in which the angle is less than 60°, the $path$ and its successor are just appended to $B$ array. Furthermore, there is one case in which the algorithm cannot determine whether there was any possible turning event. It happens when the $Previous_{road\ direction}$ is equal to $null$. Thus, the angle and the turn direction cannot be calculated. Hence, the successor of the $path$ is considered as a straight continuation of the $path$. In addition, the algorithm uses the $From\ Turn$ attribute in order to determine whether it is the main algorithm call or it is a recursion call. If $From\ Turn$ is equal to True, it means that this is the first call of the algorithm, which means that if the angle is between 60° - 180° , and the $Direction\ of\ Turn$ is equal to turning direction between the previous and current road, the successor is just a straight continuation of the $path$ and is not considered as a possible turn. In all other cases, the new path which consist of the $path$ and its successor are marked as "closed" and appended to $B$ array. The algorithm appends the $path$ and its successor which is not marked as "closed", only if the distance of the $path$ and its successor is less than $W(Path, Successors[i]) \cdot \Delta t$ which is maximum distance that can be reached.

In steps 35-36 we partition the paths that are stored in $B$ array into 2 arrays. The first array $A$ would store only the $h$ most popular opened paths, while $B$ array would store only closed paths. Opened paths, on a contrary to the closed paths, are paths from which the driver can proceed propagating to another vertex, according to algorithm limitations.

Finally, in steps 37-38, the algorithm recursively searches the continuation paths and eventually returns the $B$ array.

Analyzing $Straight\ Paths\ Finder$ algorithm

Let's denote $\left\lceil \frac{max_{dist}}{min_{road}} \right\rceil$ as $\alpha$. The number of hops is maximized when each road distance is equal to $min_{road}$. Thus, the number of hops that $Straight\ Path\ Finder$ algorithm reaches is bounded by $O(\alpha)$.

**Theorem 1**: The total computing time of the $Straight\ Path\ Finder$ algorithm would be bounded by $O(h^{\alpha})$.

Proof: From the 35th step, it can be examined that the $Straight\ Path\ Finder$ algorithm takes $h$ most popular paths and proceeds to the next step which in turn goes recursively to the successors of the last vertex of the $path$. If the algorithm produces maximum number of hops, the total computing time is equal to $\sum_{m=1}^{\alpha} h^m = h + h^2 + h^3 + \cdots + h^{\alpha} = O(h^{\alpha})$.

### C. Defining *Finding Paths* algorithm

The $Finding\ Paths$ algorithm, gets several parameters from $Getting\ Popular\ Paths$ algorithm and finds all possible paths from a last vertex in the $path$ either until number of turns in every path is equal to predefined criteria, or the path reached a dead-end or the number of turns is equal to number of rows in $MC$ file. A dead-end state is a state of the graph $G$ when the algorithm cannot reach any other vertex from a specific vertex under specific constraints. The predefined criteria is a parameter $m$ which is the maximum turning events in all paths that $Finding\ Paths$ algorithm returns to $Getting\ Popular\ Paths$ algorithm. In the input there are several attributes that assist $Finding\ Paths$ algorithm on finding continuation to the current $path$. The algorithm gets 2 attributes, $Triptime$ and $MC$ file, that allow it to estimate the time boundaries in which it has to find the successor paths to the last vertex of the $Path$. In addition, it gets parameter $m$ which notates the maximum number of turning events that the $FindingPaths$ algorithm can reach. The output of the algorithm returns possible continuation paths from the last vertex of the $Path$.

---

$\boldsymbol{Finding\ Paths}\ (Triptime, Path, MC, m)$

1) $B = \{\}$
2) $\Delta t = |MC[turn_num_{Path}][1] - TripTime|$
3) $node_{successors}$=$Straight_{Path_{Finder}}$
4) $(\Delta t, MC[turn_{num_{Path}}][0], [Path[Path.length - 1]], True)$
5) $\boldsymbol{for}\ path\ in\ node_{successors}$:
6) $delete\ the\ last\ node\ and\ the\ "closed"\ mark$
7) $\boldsymbol{If}\ length(node_{successors}) == 0$:
8) $Mark\ Path\ as\ "closed"$.
9) $B.append_one_path(Path)$
10) $\boldsymbol{return}\ B$
11) $\boldsymbol{else}$:
12) $\boldsymbol{for}\ i \leftarrow 1\ to\ i \leftarrow length(node_{successors})$:
13) $B.append_{paths}(Path.add_paths(node_{successors}[i]))$
14) $\boldsymbol{If}\ turn_num_{B[i]} == Length[MC]$:
15) $Mark\ B[i]\ as\ "closed"$
16) $A = B$
17) $\boldsymbol{for}\ i \leftarrow 1\ \boldsymbol{to}\ i \leftarrow Length[A]$:
18) $WH = False$
19) $\boldsymbol{If}\ turn_num_{A[i]} < m\ and\ A[i]\ is\ not\ marked\ as\ "closed"$:
20) $n = out\ vertex\ of\ the\ last\ edge\ in\ A[i]$
21) $\boldsymbol{for}\ j \leftarrow 1\ \boldsymbol{to}\ j \leftarrow Length[B]$:
22) $\boldsymbol{If}\ B[j] \neq A[i]$ && $n$ equals to one of turning nodes in $B[j]$ && $n\ != B[j][length(B[j]) - 1]$:
23) $WH = True$
24) $k = position\ of\ successor\ node\ of\ n\ in\ B[j]$
25) $P = []$
26) $\boldsymbol{while}\ turn_num_P \leq length(MC)\ or\ turn_num_P \leq m$:
27) $P.append_vertex(B[j][k])$
28) $k$++
29) $B.append_one_path(A[i].add_paths(P))$
30) $A.append_one_path(A[i].add_paths(P))$
31) $\boldsymbol{If}\ A[i] \in B$:
32) $delete\ A[i]\ from\ B$
33) $\boldsymbol{If}\ WH == False$:
34) $T$= MC[$turn_num_{A[i]}$+1][1]
35) $B.append_{paths}$
36) $(Finding\ Paths(\ T, A[i], MC, m))$
37) $\boldsymbol{If}\ A[i] \in B$:
38) $delete\ A[i]\ from\ B$
39) $\boldsymbol{return}\ B$

---

$Finding\ Paths$ algorithm logic

The Algorithm uses two arrays that hold the paths. Array $A$ holds all the potential paths that have not been completed enough turning events according to $MC$ file, while Array $B$ holds all possible paths. During each iteration, the algorithm goes recursively throughout possible paths which are limited in time, turning direction and average maximum legal speed. If there are no continuation paths for the $Path$, the $Path$ is marked as "closed" and the algorithm returns the $Path$ with "closed" mark to higher level recursion. In order to bring down the computation time of the algorithm, before going to another recursion, the algorithm checks whether the continuation of the $Path$ already exists in other paths which were examined before. In that case, the algorithm copies the rest of the path to $A, B$ arrays, with the restrictions that we mentioned before.

$Finding\ Paths$ algorithm steps

In step number 1, the $Finding\ Paths$ algorithm defines $B$ array, which later on will encapsulate all of our possible paths. In the 3rd step, the algorithm retrieves, using the $Straight\ Path\ Finder$ algorithm, all of the possible continuation paths with a time boundary of $\Delta t$ . Therefore, in steps 6-11 if there are no consequent trajectories, the algorithm marks the input $path$ as "closed" and returns it. Otherwise, in steps 12-15, if the input $path$ with its continuation path encapsulates the same number of turning events as the length of $MC$ file, it is marked as "closed". Therefore, we introduce auxiliary array $A$ which in the 16th step holds the paths of $B$ array. The $A$ array would hold all of the $B$ array paths. The role of $A$ array is to keep the order of path search. Steps 17-36 constitute the main core of the algorithm in which it iterates through all of the paths in array $A$ and finds recursively, all of the possible paths which are bounded by time, turning direction, legal speed and maximum turning events $m$.

In order to reduce the computation time of the algorithm, before going to another recursion, it checks, in steps 21-32, whether the algorithm has already visited the current vertex. In this case, it copies the continuation path from a previous trajectory that is already containing the continuation path of the current vertex. The copying of the continuation path is within the boundaries of $MC$ file size. Finally, in steps 33-36, in case that the algorithm didn't manage to find the continuation path in other paths, it recursively calls for $Finding\ Paths$ algorithm.

Analyzing $Finding\ Paths$ algorithm

If a distance $X_{turn}$ is equal to $min_{road}$, then the number of possible turns that algorithm $Straight\ Path\ Finder$ would produce is bounded by $\left\lfloor\frac{max_{dist}}{X_{turn}}\right\rfloor = \left\lfloor\frac{max_{dist}}{min_{road}}\right\rfloor = O(\alpha)$ turns. Thus, the number of possible turns which the $Straight\ Path\ Finder$ algorithm would find is bounded by $O(\alpha)$.

**Lemma 2** – The computation time of the $PathFinding$ algorithm without checking previous paths equals to $O(h^{\alpha} \cdot \alpha^{m-1})$.

Proof: The number of possible turns which the $Straight\ Path\ Finder$ algorithm would find is bounded by $O(\alpha)$. In every recursion, the $PathFinding$ algorithm reveals $\alpha$ new paths while calling $Straight\ \ Path\ Finder$ which computation time is bounded by $O(h^{\alpha})$. Thus, the computation time would be $h^{\alpha} + \alpha \cdot\ h^{\alpha} + \alpha^2 \cdot\ h^{\alpha} + \cdots + \alpha^{m-1} \cdot\ h^{\alpha} =\ h^{\alpha} \cdot (1 + \alpha + \alpha^2 + \alpha^3 + \cdots + \alpha^{m-1}) = O(\ h^{\alpha} \cdot \alpha^{m-1})$.

**Lemma 3** – The computation time for iterating in previous paths is equal to $O(m \cdot \alpha^{2m})$.

Proof: In the worst case scenario, the $Finding\ Paths$ algorithm would iterate through all of the possible paths until it will reach $m$ turning events in all the paths. Thus, in every turning vertex, it will have to iterate through all the possible previous paths in $B$ array. The maximum number of possible paths is bounded by $\alpha^m$ . For each possible path, the algorithm will iterate through $m \cdot \alpha^m$ vertices, which is the maximum number of vertices in the $B$ array. Thus the maximum computation time for checking previous paths in steps 16-36 of $Finding\ Paths$ algorithm, is bounded by $O(m \cdot \alpha^m \cdot \alpha^m) = O(m \cdot \alpha^{2m})$.

**Theorem 4** – The total computation time of $Finding\ Paths$ algorithm equals to $O(\ h^{\alpha} \cdot \alpha^{m-1} + m \cdot \alpha^{2m})$.

Proof: The theorem complies from lemma 2 and lemma 3.

### D. Defining $\boldsymbol{Getting\ Popular\ Paths}$ algorithm

The $Getting\ Popular\ Paths$ algorithm finds all possible paths, while after each call to the $Finding\ Paths$ algorithm, it deletes all of dead-end paths.

The $Getting\ Popular\ Paths$ algorithm gets the $MC$ file which holds the cornering event time and $starting\ vertex$, which is an intersection from where the driver started his path. In addition, it gets $m$ parameter which determines the maximum number of new turning events within the paths which the $Finding\ Paths$ algorithm should return. Ranked paths according to the average popularity of each path are returned/

---

$\boldsymbol{Getting\ Popular\ Paths\ algorithm}$ ($\boldsymbol{Starting\ ve}$

1) $c = 1$
2) $Paths = FindingPaths(00{:}00, [Starting\ vertex], MC, m \cdot c)$
3) $c++$
4) $\boldsymbol{while}\ there\ is\ an\ opened\ path\ in\ Paths$:
5) $\boldsymbol{for}\ j \leftarrow 1\ to\ j \leftarrow Length(Paths)$:
6) $\boldsymbol{If}\ Paths[j]\ is\ marked\ as\ \text{“}closed\text{”}$ && $turn_num_{Paths[j]} < Length(MC)$:
7) $delete\ Paths[j]$
8) $\boldsymbol{for}\ j \leftarrow 1\ to\ j \leftarrow Length(Paths)$:
9) $\boldsymbol{If}\ Paths[j]\ is\ not\ marked\ as\ closed$:
10) $T =\ MC[turn_num_{Paths[j]} + 1][1]$
11) $Paths = Paths \cup FindingPaths(T, Paths[j], MC, m \cdot c)$
12) $delete\ Paths[j]$
13) $c++$
14) $\boldsymbol{return}\ sorted\ Paths\ array\ by\ average\ popularity$

---

$Getting\ Popular\ Paths$ algorithm, by using $Finding\ Paths$ algorithm, finds possible paths which are available from a certain $starting\ vertex$. The aim of the algorithm is to get possible paths while deleting all of dead-end paths after each iteration of $Finding\ Paths$ algorithm. Consequently, in the output it would provide paths which are ranked by average popularity.

$Getting\ Popular\ Paths$ algorithm steps

In the 2nd step, the algorithm uses $Finding\ Paths$ algorithm in order to find possible paths from $starting\ vertex$. Consequently, in steps 5-7, the algorithm checks whether a path has a dead-end. The indication for it, is when $Finding\ Paths$ algorithm marked the path as “closed”, while the current number of turns has not reached the number of turns that can be concluded from $MC$ file. Afterwards, in steps 8-12, the $Getting\ Popular\ Paths$ algorithm, for every remaining path, looks for continuation paths. The $Getting\ Popular\ Paths$ algorithm calls $Finding\ Paths$ algorithm and requires from it to retrieve all of the continuation paths until it reaches at maximum $m \cdot c$ turning events. Thus, iterating steps 5-13 until the number of turns in every path would comply with number of turns defined in $MC$ file. In other words, when the path turns reach the number of rows in $MC$ file, the $Finding\ Paths$ algorithm, marks it as "closed".

**Lemma 5** – The overall number of iterations through 5-13 steps is bounded by $\left\lceil \frac{|MC|}{m} \right\rceil - 1$ .

Proof: Every time the $Finding\ Paths$ algorithm is called, it discovers at most $m$ new turning events. The $Getting\ Popular\ Paths$ algorithm will stop iterating through 5-13 steps when there will be no opened paths. The $Finding\ Paths$ algorithm will mark the path as "closed", when the number of turning events would be equal to $|MC|$. The first call to $Finding\ Paths$ is made in 2nd step and other calls are made within steps number 5-13. Thus, the overall number of iterations through 5-13 steps is bounded by $\left\lceil \frac{|MC|}{m} \right\rceil - 1$.

**Lemma 6** – The number of paths which $Finding\ Paths$ algorithm finds is bounded by $O\left(\alpha^{|MC|}\right)$.

Proof: $Getting\ Popular\ Paths$ algorithm in 2nd step, calls for $Finding\ Paths$ algorithm, the former finds at maximum $\alpha^{\ m}$ paths for the $Starting\ vertex$. Afterwards, $Getting\ Popular\ Paths$ algorithm, iterates through steps number 5-13 until all of the paths are marked as "closed". In each iteration, the algorithm calls for $Finding\ Paths$ algorithm, while the former finds at most $\alpha^{\ m}$ new paths for last vertex for each of the paths in $Paths$ array.

Let's examine the number of paths after each iteration:

| Iter. # | Start | 1 | $i$ | $\left\lceil \frac{\lvert MC \rvert}{m} \right\rceil - 2$ | $\left\lceil \frac{\lvert MC \rvert}{m} \right\rceil - 1$ |
|---|---|---|---|---|---|
| #paths | $\alpha^{\ m}$ | $\alpha^{\ 2m}$ | $\alpha^{\ (i+1)m}$ | $\alpha^{(\lceil \frac{\lvert MC \rvert}{m} \rceil - 1)\cdot m}$ | $\alpha^{\lceil \frac{\lvert MC \rvert}{m} \rceil \cdot m}$ |

Thus, number of paths is bounded by $\alpha^{\left\lceil \frac{|MC|}{m} \right\rceil \cdot m} = O\left(\alpha^{|MC|}\right)$ .

**Theorem 7** – Overall computation time of $Getting\ Popular\ Paths$ algorithm is bounded by $O\left(\left(\alpha^{\left(\left\lceil \frac{|MC|}{m} \right\rceil - 1\right)\cdot m} + 1\right) \cdot \{\, h^{\alpha} \cdot \alpha^{m-1} + m \cdot \alpha^{2m}\}\right)$ .

Proof: According to Theorem 4, the computation time for each call to $Finding\ Paths$ is bounded by $O(\, h^{\alpha} \cdot \alpha^{m-1} + m \cdot \alpha^{2m})$. In the 2nd step, the $Getting\ Popular\ Paths$ algorithm calls for $Finding\ Paths$ algorithm and finds at most $\alpha^{m}$ paths for the $Starting\ vertex$. Before steps number 8-12, there will be the following number of paths in $Paths$ array, which equals to number of calls to $Finding\ Paths$ algorithm:

| It. # | 1 | 2 | $i$ | $\left\lceil \frac{\lvert MC \rvert}{m} \right\rceil - 2$ | $\left\lceil \frac{\lvert MC \rvert}{m} \right\rceil - 1$ |
|---|---|---|---|---|---|
| #paths before steps 8-12 | $\alpha^{\ m}$ | $\alpha^{\ 2m}$ | $\alpha^{\ i\cdot m}$ | $\alpha^{(\lceil \frac{\lvert MC \rvert}{m} \rceil - 2)\cdot m}$ | $\alpha^{\left(\lceil \frac{\lvert MC \rvert}{m} \rceil - 1\right)\cdot m}$ |

Thus, the overall computation time is a combination of computation time in the 2nd step which is bounded by $O(\, h^{\alpha} \cdot \alpha^{m-1} + m \cdot \alpha^{2m})$ and all of the calls to $Finding\ Paths$ algorithm in each iteration which is bounded by $O\left(\sum_{c=1}^{\left\lceil \frac{|MC|}{m} \right\rceil - 1} \alpha^{cm} \cdot \{\, h^{\alpha} \cdot \alpha^{m-1} + m \cdot \alpha^{2m}\}\right)$. Thus, the overall computation time for all calls is as follows:

$$\{\, h^{\alpha} \cdot \alpha^{m-1} + m \cdot \alpha^{2m}\} + \sum_{c=1}^{\left\lceil \frac{|MC|}{m} \right\rceil - 1} \alpha^{cm} \cdot \{\, h^{\alpha} \cdot \alpha^{m-1} + m \cdot \alpha^{2m}\} = \{\, h^{\alpha} \cdot \alpha^{m-1} + m \cdot \alpha^{2m}\} + \alpha^{m} \cdot \{\, h^{\alpha} \cdot \alpha^{m-1} + m \cdot \alpha^{2m}\} + \cdots + \alpha^{\left(\left\lceil \frac{|MC|}{m} \right\rceil - 1\right)\cdot m} \cdot \{\, h^{\alpha} \cdot \alpha^{m-1} + m \cdot \alpha^{2m}\} = O\left(\left(\alpha^{\left(\left\lceil \frac{|MC|}{m} \right\rceil - 1\right)\cdot m} + 1\right) \cdot \{\, h^{\alpha} \cdot \alpha^{m-1} + m \cdot \alpha^{2m}\}\right).$$

### E. $\boldsymbol{Retrieving\ Driver's\ Paths}$ algorithm

Finally, we are using all of previous components for the following algorithm In the input of the algorithm there is $G = (V, E)$ directed graph which represents the area of driving, while the vertices represent intersections or turns greater than 60° and edges represent the roads. Furthermore, the algorithm gets the $starting\ vertex$, $S_{average}$ which is driver's average speed, $MC$ file, and $\Delta t_{\ total}$ which is the total driving time. Furthermore, the algorithm gets parameter $m$ which defines number of new turning events for each call for $Finding\ Paths$ algorithm. Ordered array of all possible paths, for driver movement serves as an output for our algorithm. The possible paths are ordered by the average road popularity.

| $\boldsymbol{Retrieving\ Driver's\ Paths}$ |
|---|
| 1. $G = Getting\ Popularities\ ()$ |
| 2. $Paths = Getting\ Popular\ Paths$ $(Starting\ vertex, MC, m)$ |
| 3. $\boldsymbol{for}\ path\ in\ Paths$: |
| 4. $\quad \boldsymbol{If}\ S_{average} * \Delta t_{\ total} * 1.1 <$ |

$|path| \;\; || \;\; |path| < S_{average} * \Delta t_{\;total} * 0.9$:

5. $delete\ (path)$

6. $\boldsymbol{return}\ Paths$

In step 1, the algorithm is mapping the popularities for each edge by using $Getting\ Popularities$ algorithm. Consequently, in step 2, the algorithm gets possible paths that can be reached from a $Starting\ vertex$ and bounded by $MC$ file, while $\Delta t_{total}$ is the total trip time of the driver. Afterwards in steps 3-5, it deletes all of the paths that do not meet the user's average speed attribute. Finally, the algorithm returns all of the remaining paths which are ranked by average popularity of the path.

**Theorem 8** – The total computation time of $Retrieving\ Driver's\ Paths$ algorithm is bounded by

$$O\left(|E| \cdot |C| \cdot |N| + \left(\left(\alpha^{\left(\left\lceil\frac{|MC|}{m}\right\rceil-1\right)\cdot m} + 1\right) \cdot \{\, h^{\alpha} \cdot \alpha^{m-1} + m \cdot \alpha^{2m}\}\right)\right).$$

Proof: The computation time derives from computation time of $Getting\ Popularities$ algorithm and Theorem 7.

## V. SIMULATION

This section describes the results of the path reconstruction algorithms which were implemented on the real vehicle's paths. A description of the set-up is followed by a presentation of the findings and their analysis.

### A. Environment Set-up

In order to test our algorithms we used the dataset of real life trajectories. The dataset was collected as a part of $GeoLife$ 2.0 project and was conducted and published by $Microsoft$ company. The paths were collected during 5 years by the people of Beijing [20]. In addition, in order to implement our algorithms $Python$ 2.7 coding language was used. The algorithms were tested over the trajectories with turning events in a range of 1-6 events, and trajectories distance varying from $0.337\ km$ to $8.69\ km$. The trajectories were distributed over the area of the city of Beijing.

Two types of comparative analysis were tested. First analysis compared the rank of the closest constructed path as a function of traveling distance, traveled time and average user's speed ($S_{average}$), while examining the difference among number of turning events in trajectories. Second analysis compared the maximum distance of the first ranked constructed trajectory from the real path as a function of traveled distance, traveled time and average user's speed ($S_{average}$).

In order to map all of the road popularities, we used the $Getting\ Popularities$ algorithm, while GPS log files that we used were the log files from taxi trajectories project which is called T-Drive [21]. T-Drive files contain a one-week GPS coordinates of 10,357 taxis. The total number of points is about 15 million.

We used the $Straight\ Path\ Finder$ algorithm with the following parameters: $Y$ is equal to $10km$ and $h = 2$. The reason for our selection is that the computation times for those parameters are reasonable while constructing good trajectories.

We used the $Retrieving\ Driver's\ Paths$ algorithm with the $m = 3$ parameter. The reason for our choice is our will to use $m$ that will be high enough, so we will not lose essential paths when deleting non popular paths in $Getting\ Popular\ Paths$ algorithm, and low enough for obtaining computation time.

### B. $\boldsymbol{Retrieving\ Driver's\ Paths}$ Performance

We now present the findings for $Retrieving\ Driver's\ Paths$ algorithm, while examining the influence of the traveling distance and the average user speed $S_{average}$ and traveling time on the closest constructed path to the real trajectory. We use the notion of standings as a popularity assessment of each path in the $Paths$ array, while the most popular path will be ranked as 1 and the least popular path will be ranked as the length of the $Paths$ array.

Traveled Distance Influence

In Fig. 5.1 we present the traveled distance influence on the absolute standings, respectively, of the closest constructed path to the real user's path. We examined 50 trajectories which traveling distances variant from $0.337\ km$ to $8.69\ km$, while examining the difference between the number of rows in $MC$ file, which represents the number of turning events in the path. We notice that the absolute standing of the constructed path increases with increase of traveling distance. Furthermore, one can see that the number of turning events makes no difference on the absolute standings of the constructed path. The absolute standings of the path is beginning to have a bigger variance after traveling distance of $4\ km$ and its very high as we reach the $8 - 9\ km$. The conclusion is that we can reveal user's path with an absolute standings up to 20 as long as his traveled distance is lower than $4\ km$. In addition, we cannot see any influence of the traveled distance on the relative standings.

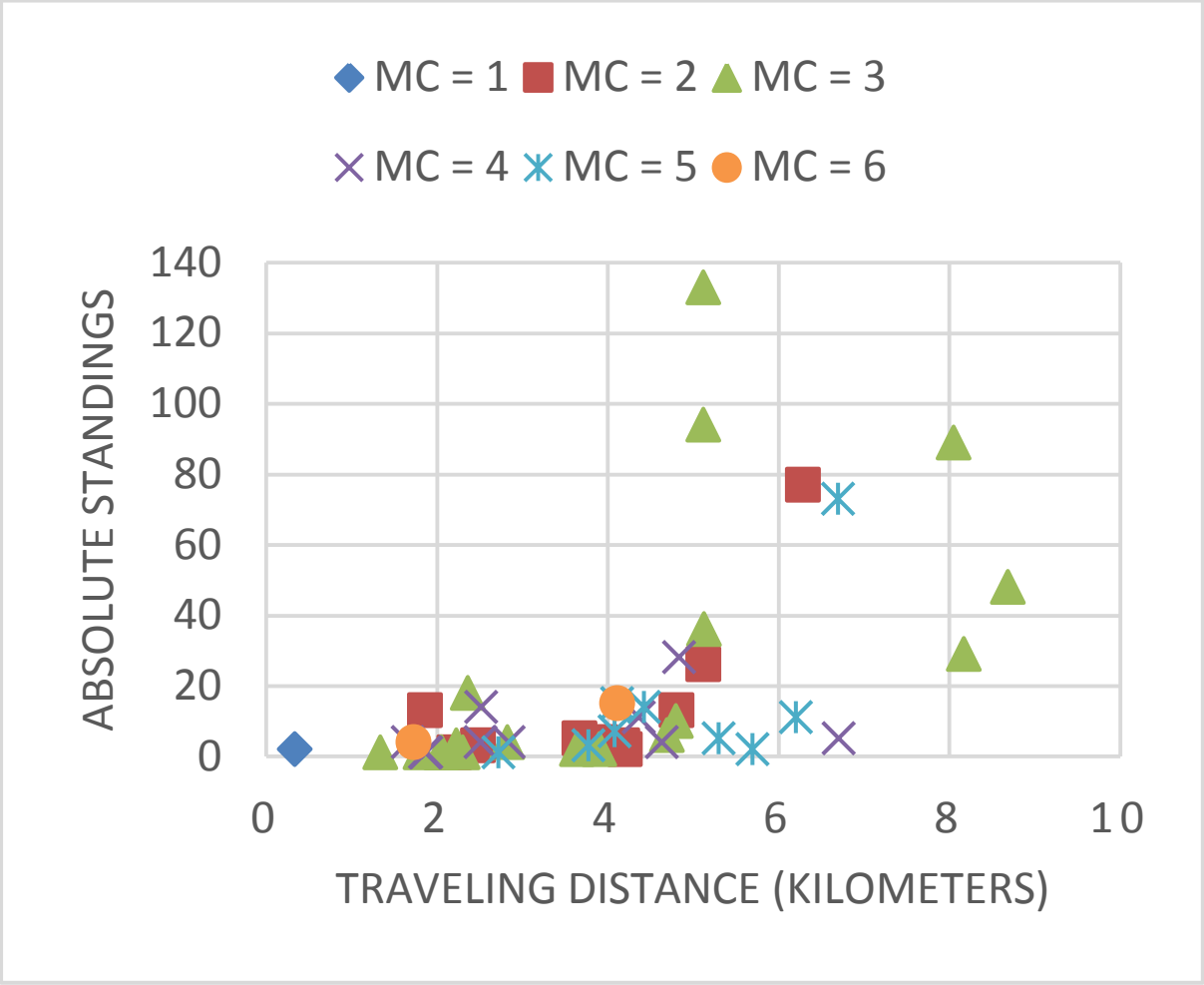

**Figure 5.1: Absolute standings of the constructed path as a function of traveling distance**

Average Speed ($S_{average}$) Influence

In Fig. 5.2 we present the user's average speed influence on the absolute standings of the closest constructed path to the real user's path. We examined 50 trajectories which speed variant from 7.46 $km/h$ to 40.96 $km/h$. From figure 5.2, it seems that the average speed doesn't have any effect on the absolute standings of the constructed path. Furthermore, number of turns also doesn't influence the absolute standings of the constructed path.

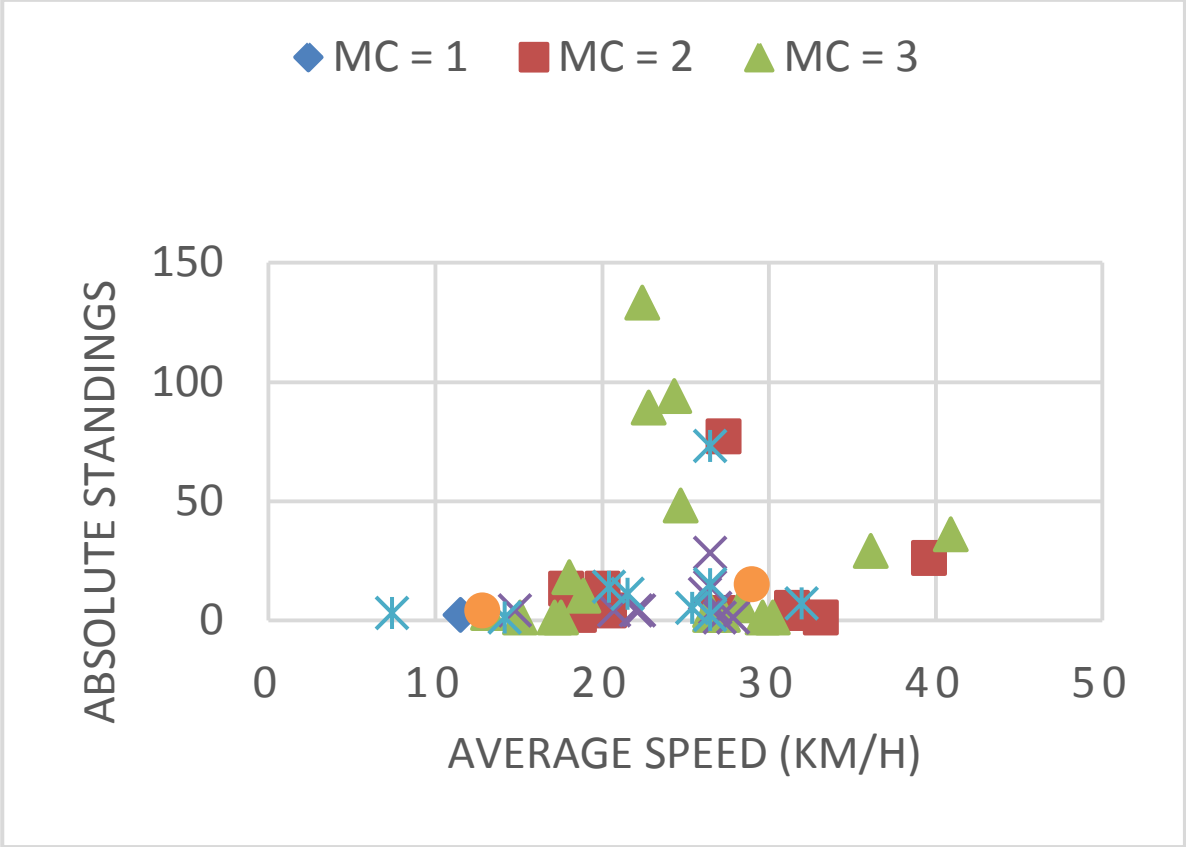

**Figure 5.2: Absolute standings of the closest constructed path as a function of user's average speed.**

Influence of Traveling Time

In Fig. 5.3 we present the user's traveling time influence on the absolute standings of the closest constructed path to the real user's path. We examined 50 trajectories while traveling time varies from 1.75 minutes to 30.4 minutes.

As expected, we notice that the absolute standing of the constructed path increases with the increase of traveling time. Furthermore, one can see that number of turning events makes no difference on the absolute standing of the constructed path. The absolute standings is starting to have a bigger variance after traveling time of 7 minutes and is getting very high as we reach 12.5 minutes. The conclusion is that we can reveal user's path as long as his traveling distance is lower than 7 minutes, while the constructed path has absolute standings is in top 20 trajectories.

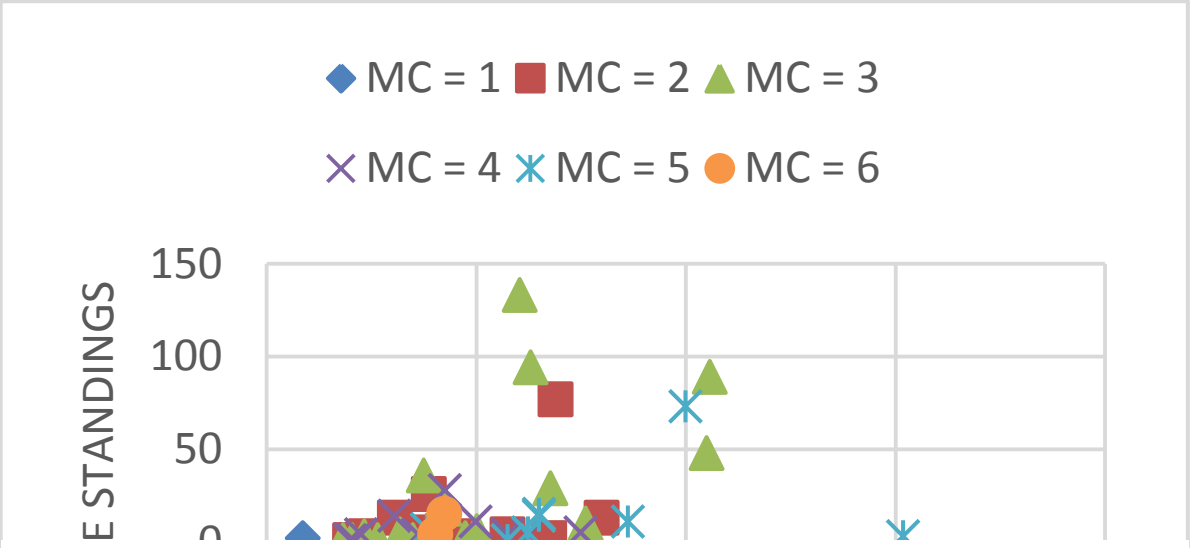

**Figure 5.3: Absolute standings of the constructed path as a function of user's traveling time**

Maximum distance from trajectory

In order to examine the maximum distance between the constructed trajectory and the driver's path, we tested the maximum distance from the first ranked constructed trajectory. Three types of comparative analysis were performed. First analysis compared the maximum distance of the first ranked constructed path as a function of traveled distance. Second analysis compared the maximum distance of the first ranked constructed path as a function of user's average speed ($S_{average}$). Third analysis compared the maximum distance of the first ranked constructed path as a function of traveling time.

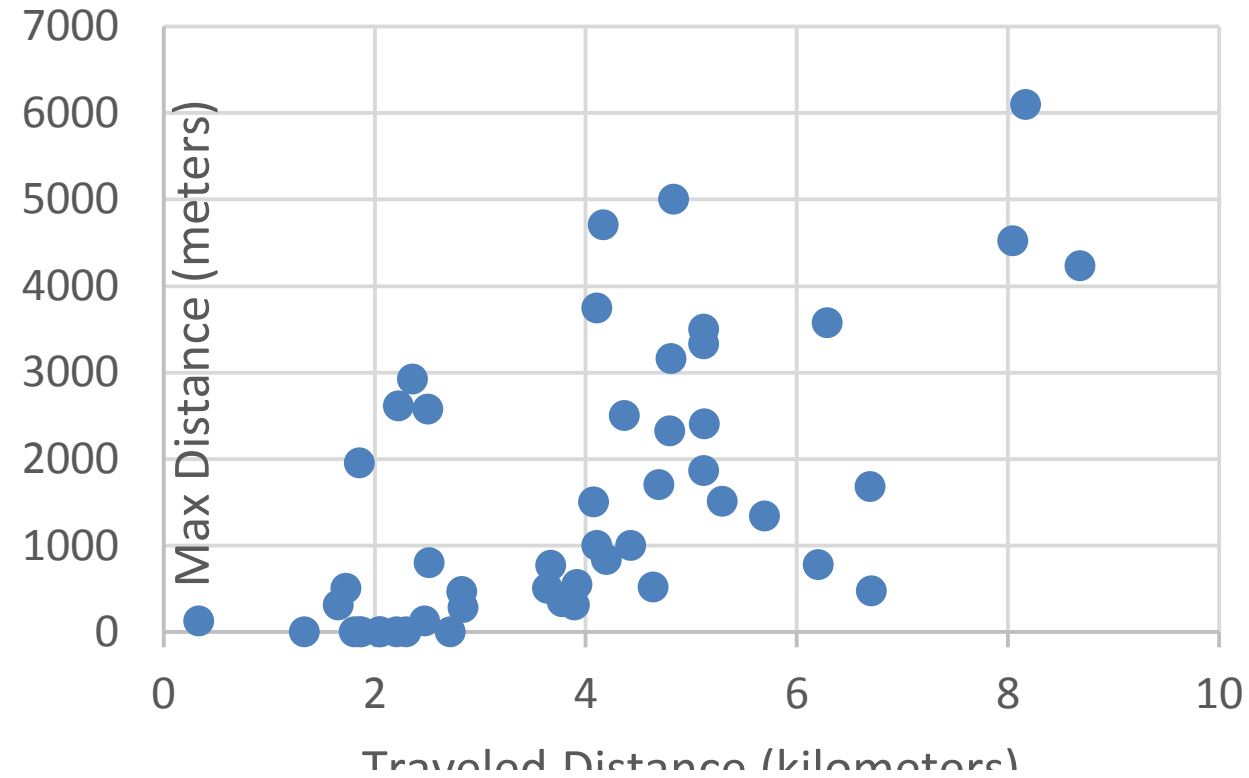

**Figure 5.4: Maximum distance of the first ranked constructed path as a function of traveled distance.**

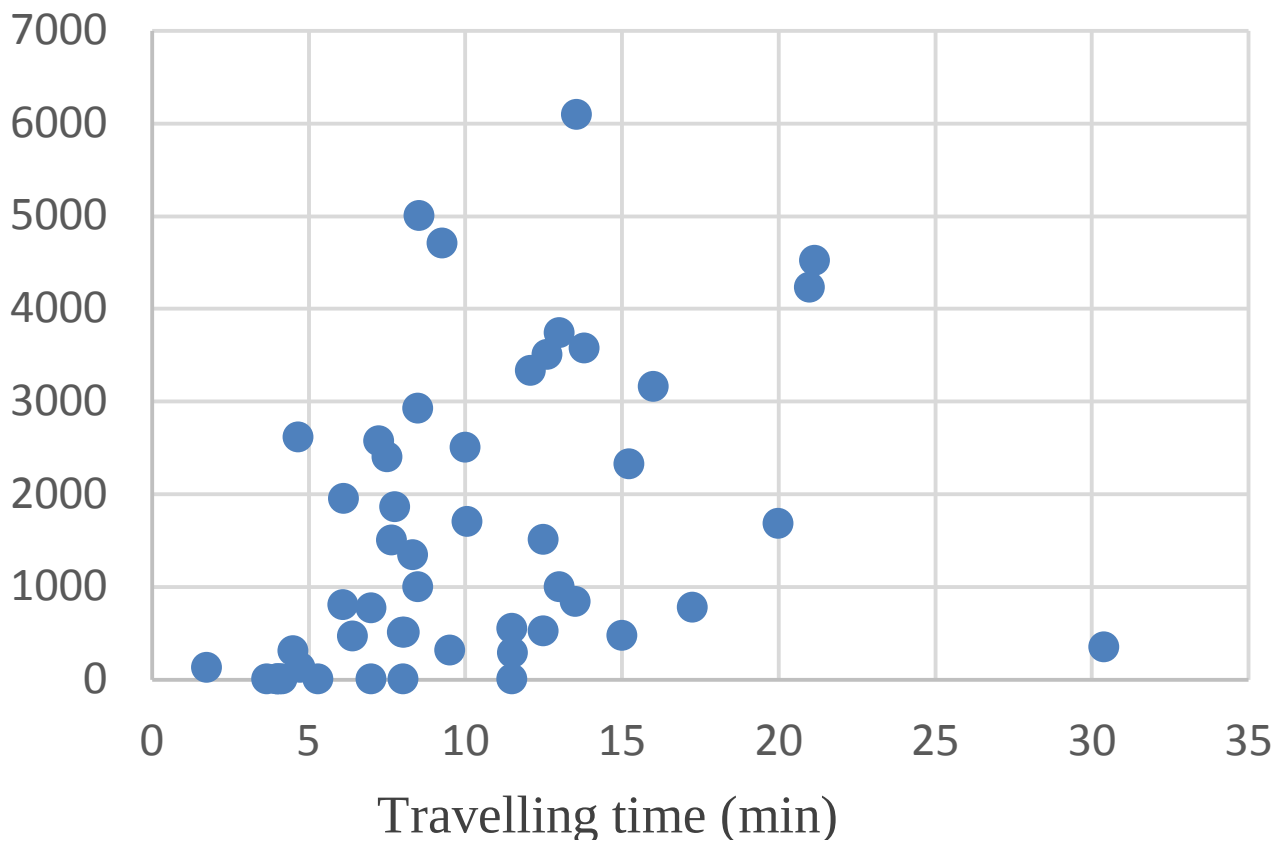

**Figure 5.5: Maximum distance of the first ranked constructed path as a function of traveling time.**

From Fig. 5.4 one can learn that as traveling distance is getting higher, the maximum distance is getting higher with greater variance. The ratio between the axes has a proportion lower than 1 in a majority of trajectories. Furthermore, number trajectories which have maximum distance lower than 10%, 15%, 20% of the driver's traveled distance are equal to 12, 15, 19 out of 50 trajectories, respectively. From Fig. 5.5, we observe that the traveling time has no influence on the maximum distance. From Fig. 5.6, we observe that average speed has the same influence on maximum speed as the influence of traveling distance.

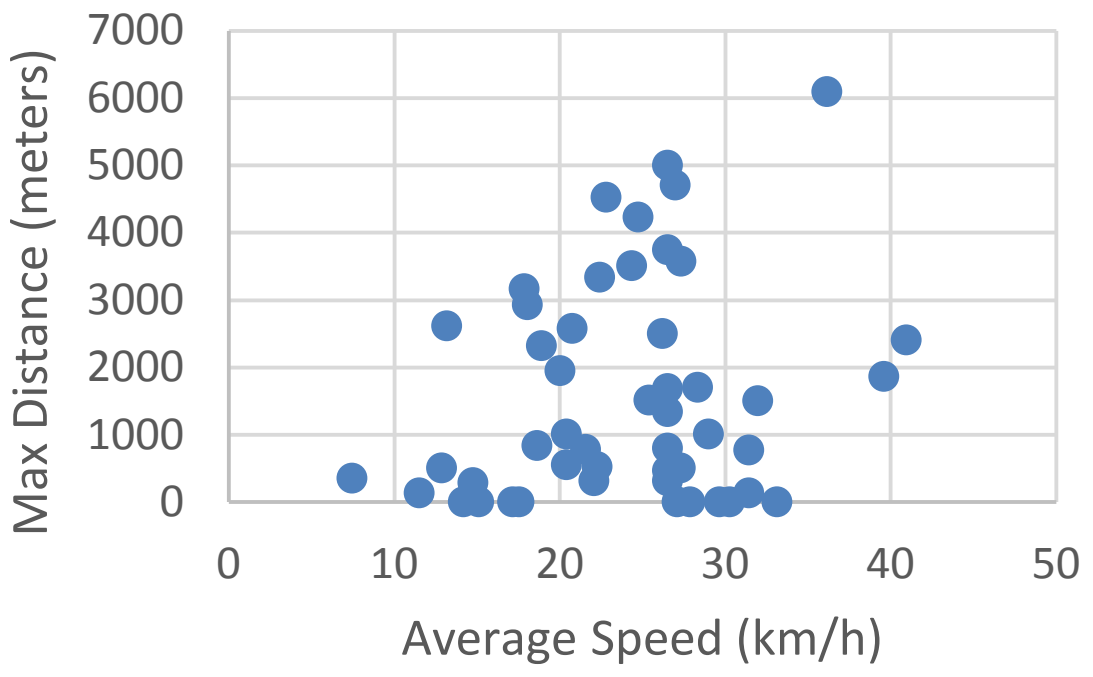

**Figure 5.6: Maximum distance of the first ranked constructed path as a function of average speed.**

<u>Types of trajectories and their influence on maximum distance</u>

There are 3 types of trajectories which we examined. First type of trajectory is a path in which the driver starts from residential road and finish his path in the highway. Second type of a trajectory is a path in which the driver starts from a residential area, propagates via highway roads and finishes his path in residential area. Third type is a trajectory in which the driver starts from the highway.

In order to understand the difference between the roads and their influence of the roads popularity, we compared the maximum distance between the real trajectory and the most popular (1st place), random and median ranked trajectory.

*a) Residential to highway trajectory*

We can notice that in 7 out of 7 paths, the 1st ranked path was closer to the real trajectory than a median ranked path, see Fig. 5.7. Thus, the popularity plays an important role in that kind of trajectories.

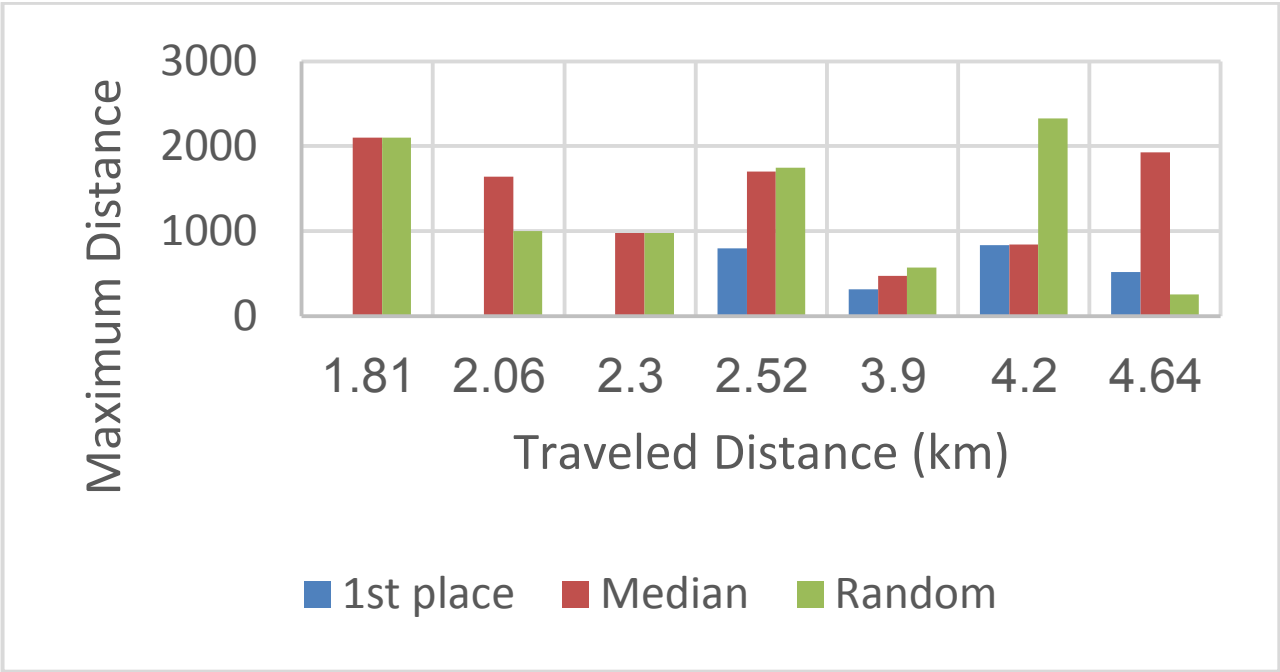

**Figure 5.7: Maximum distance of the first ranked, median and random trajectory as a function of traveled distance in paths which start in the residential road and finish in the highway.**

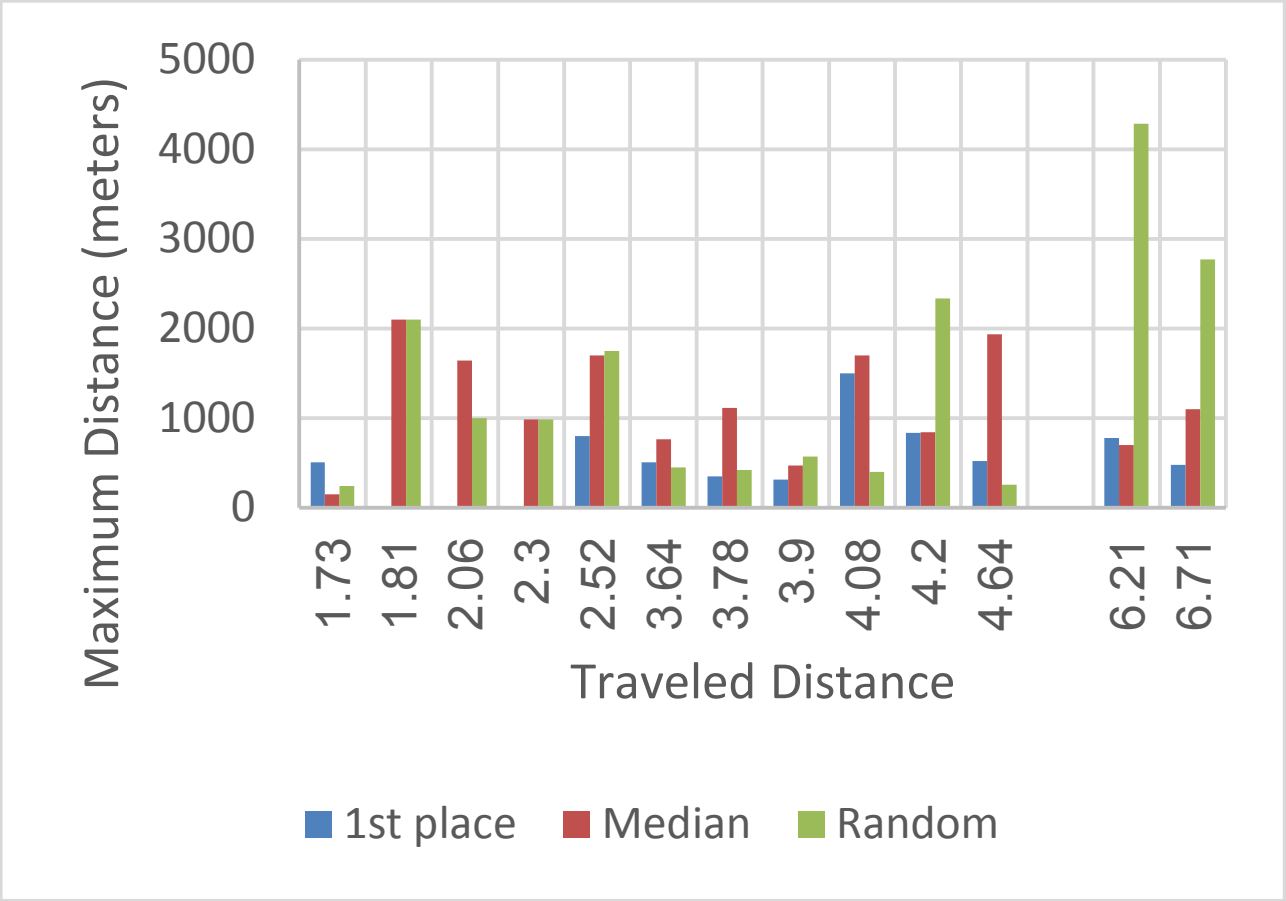

**Figure 5.8: Maximum distance of the first ranked, median and random trajectory as a function of traveled distance in paths which start in the residential road and finish in the highway and roads which start from the residential road, propagates through highway and finish in residential road.**

*b) Residential through highway to residential trajectories*

The setup is:

1. The driver started from residential road and ended his path in highway.
2. The driver started from residential road, drove through the highway and ended his path in residential area.

We can notice (Fig. 5.8) that in 12 out of 14 trajectories, the 1st ranked path was closer to the real trajectory than a median ranked path. On the one hand, in the majority of trajectories, the 1st ranked path was closer to the real trajectory than a median, which means that the popularity has a major influence on the constructed path. On the other hand, when we added a second type of trajectories to our bar chart, we can notice that not all of the 1st place trajectories are closer than a median ranked paths to the real trajectory, which means that the second type of trajectories add a certain uncertainty.

*c) Trajectories which start from the highway*

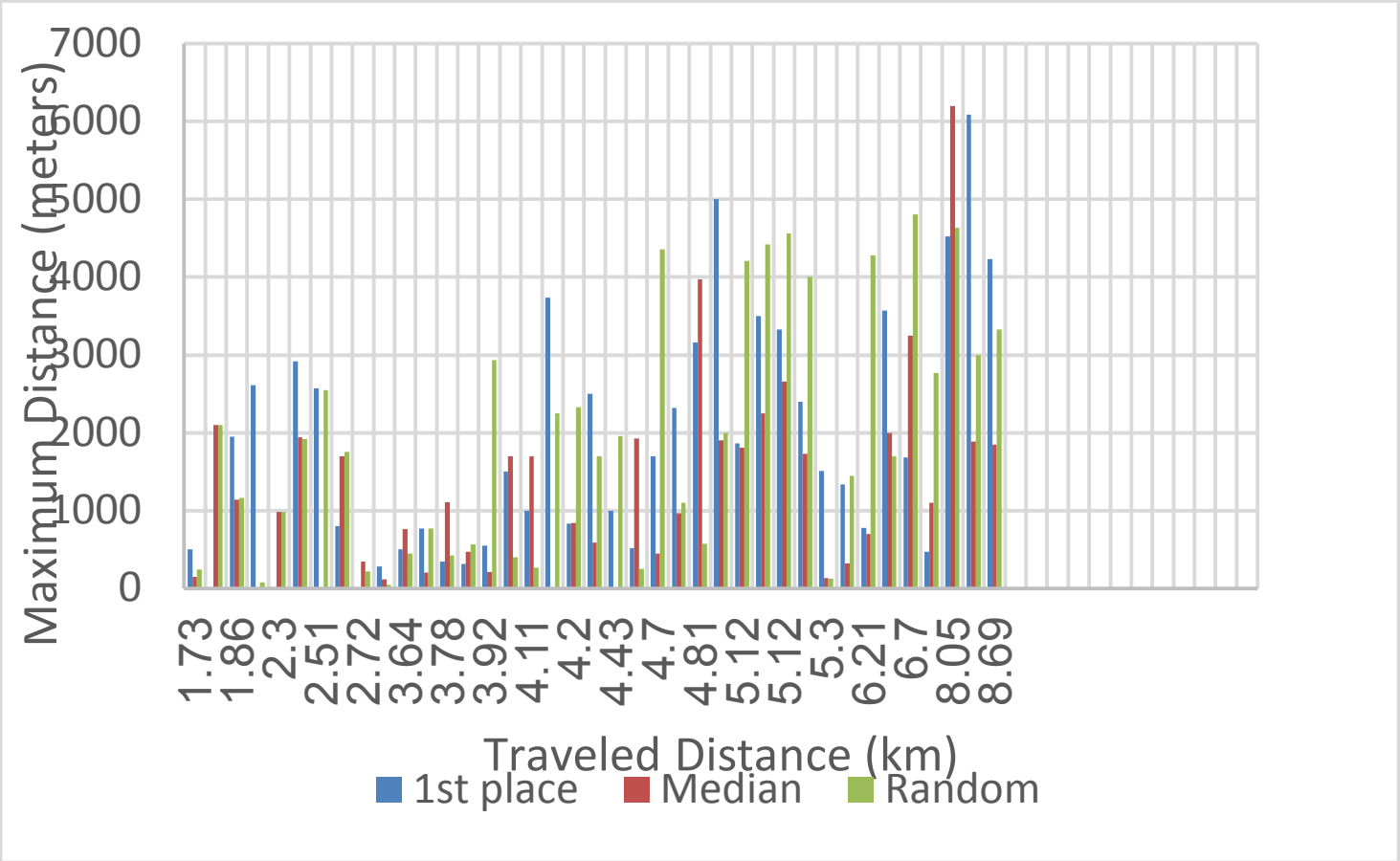

**Figure 5.9: Maximum distance of the first ranked, median and random trajectory as a function of traveled distance in all types of path**

Figure 5.9 concludes all 3 types of the trajectories. We can notice that by adding the $3^{rd}$ types of trajectories (trajectories which start from a highway), the majority of trajectories which are closer to the real trajectories are median trajectories. Thus, in a $3^{rd}$ type of trajectories, the popularity doesn't play an important role for constructing driver's trajectory.

### C. General Findings

There are several findings that we noticed while running our algorithms. First, when there is an interchange in driver's path, our algorithms produce constructed path which doesn't have a good absolute standings, with no difference what is the traveling time, traveling distance or average speed of the user. Second, because of a high amount of possible paths, our algorithms are performing in a reasonable computation time when a path has a distance up to 10 $km$ . Third, motorway and freeway roads have higher popularity than residential roads. Thus, when the driver drives on the residential road, which is located near the motorway or freeway, the constructed path would be ranked in worse position. Finally, there are some situations in which the real path is passing through private roads. Hence, there will be some difficulty to reconstruct his path. The constructed path might be close or far from the original road, depends on the case. In addition, there are some paths which we ignored since the popularities of those paths were too low. The amount of that kind of paths was about 7% of the total amount of paths that we examined.

## VI. CONCLUSIONS AND FUTURE WORK

In this work we have studied the problem of breaching driver's privacy by revealing driver's path while using basic driving information. Although the problem has been studied in several articles, some of them assume that GPS coordinates of the driver in the middle and the end of the path are known, while others assume that driver's continuous speed is known. We proposed a new approach for reconstructing driver's trajectory from basic driving information and by using roads popularities. By knowing the beginning location, driver's average speed, and cornering log file, we can reconstruct driver's path while having the closest reconstructed path ranked among top ranked paths. The paths are constructible within a reasonable time.

For reconstructing the driver's path, first we found possible turning vertices from starting vertex. We continued doing so, until number of turning vertices in each path was equal to a predefined criteria. Afterwards, we deleted all of dead-end paths and continued finding continuation paths until the number of turning events was equal to the number of cornering events in cornering log file. Finally, we deleted all of the paths which are not comply with cornering log file and driver's average speed. Thus, we ranked the final paths by their popularities. When we examined the maximum distance from the first ranked constructed trajectory as a function of traveled distance, traveling time and driver's average speed, we showed that as traveled distance is getting higher, the maximum distance is getting higher with greater variance, while the same applies for user's average speed. In addition, when the path starts from residential road and finish its path in a highway, the constructed path would be ranked much higher than a path which started in highway. *This phenomenon can be explained by hypothesis that the driver tends to drive from lower to higher popularity roads, while highways have higher popularities*. In other words, the driver which starts from residential road, will tend to propagate to the highway. Furthermore, the standings of the trajectory which is closest to the real trajectory are not influenced by the amount of turning events.

The optimal solution is yet to be reached when it comes to reconstruction of driver's path while using roads popularities, and is a task for future study. Other possible directions are reconstructing driver's path while knowing GPS coordinates in the middle of the trajectory. That can be useful when the user is willing to provide his GPS coordinates but his signal is lost in some scenarios, e.g. driving in the tunnel or in areas with electro-magnetic interference.